\documentclass[a4paper,11pt]{article}
\pdfoutput=1 

\usepackage{jheppub} 

\usepackage[T1]{fontenc} 
\usepackage{mathrsfs}
\usepackage{tikz-feynman}
\usepackage{feynmf}
\usepackage{tikz}

\newcommand{\ud}{\mathrm{d}}
\newcommand{\pd}{\partial}

\newcommand{\lie}{\mathscr L}

\newcommand{\order}[1]{\mathcal O\left(#1\right)}
\newcommand{\scri}{\mathscr I^+}

\title{Gravitational memory effects and Bondi-Metzner-Sachs symmetries in scalar-tensor theories}

\author[a]{Shaoqi Hou}
\author[a,b]{Zong-Hong Zhu}

\affiliation[a]{School of Physics and Technology, Wuhan University,\\Wuhan, Hubei 430072, China}
\affiliation[b]{Department of Astronomy, Beijing Normal University,\\Beijing 100875,  China}

\emailAdd{hou.shaoqi@whu.edu.cn}
\emailAdd{zhuzh@whu.edu.cn}

\abstract{The relation between gravitational memory effects and Bondi-Metzner-Sachs symmetries of the asymptotically flat spacetimes is studied in the scalar-tensor theory.
For this purpose, the solutions to the equations of motion near the future null infinity are obtained in the generalized Bondi-Sachs coordinates with a suitable determinant condition.
It turns out that the Bondi-Metzner-Sachs group is also a semi-direct product of an infinite dimensional supertranslation group and the Lorentz group as in general relativity.
There are also degenerate vacua in both the tensor and the scalar sectors in the scalar-tensor theory.
The supertranslation relates the vacua in the tensor sector,  while in the scalar sector, it is the Lorentz transformation that transforms the vacua to each other. 
So there are the tensor memory effects similar to the ones in general relativity, and the scalar memory effect, which is new.
The evolution equations for the Bondi mass and angular momentum aspects suggest that the null energy fluxes and the angular momentum fluxes across the null infinity induce the transition among the vacua in the tensor and the scalar sectors, respectively.}

\begin{document}
\maketitle
\flushbottom

\section{Introduction}
\label{sec-int}

Gravitational memory effect is a permanent displacement of test particles after the passage of gravitational waves (GWs) discovered in general relativity (GR) a long time ago \cite{Zeldovich:1974gvh,Braginsky:1986ia,Christodoulou1991,Thorne:1992sdb}.
It is caused by the energy fluxes traveling through the (future) null infinity $\scri$ in an asymptotically flat spacetime \cite{Strominger:2014pwa}, which possesses an asymptotic symmetry group, called Bondi-Metzner-Sachs (BMS) group \cite{Bondi:1962px,Sachs:1962wk,Sachs:1962zza}.
This group is a semi-direct product of an infinite dimensional supertranslation group and the Lorentz group. 
Under the action of supertranslations, the classical vacuum is degenerate. 
The energy fluxes induce the transition among the degenerate vacua and result in the memory effect.
Memory effect gains an increasing interest due to the study of the infrared structure of gravity \cite{Strominger:2014pwa,Strominger2014bms,Pasterski:2015tva,Strominger:2018inf} and the detection of GWs \cite{Abbott:2016blz,Abbott:2016nmj,Abbott:2017vtc,Abbott:2017oio,TheLIGOScientific:2017qsa,Abbott:2017gyy,LIGOScientific:2018mvr,Abbott:2020uma}.
The former reveals the infrared triangle -- the equivalence among memory effect, soft theorem \cite{Weinberg:1965s} and asymptotic symmetry. 
It also stimulated the works  on the asymptotic symmetry on the black hole horizon \cite{Hawking:2016msc,Donnay2016prl,Donnay2016esh,Hawking:2016sgy,Hou:2017pes,Donnay2018bhm} and black hole information paradox \cite{Hawking1976inf}.
The later opens up the possibility to examine the existence of this effect experimentally in the near future \cite{Seto:2009nv,Wang2015mm,Lasky:2016knh,McNeill:2017uvq,Johnson:2018xly,Hubner2020mmn,Boersma:2020gxx,Madison:2020xhh}.

Alternative metric theories of gravity should also have memory effect, since they have two degrees of freedom like those in GR, at least.
In this work, we would like to study the memory effect in the scalar-tensor theory, the simplest alternative to GR.
Unfortunately, there are essentially infinitely many scalar-tensor theories, so we would like to start with the simplest one, i.e., the Brans-Dicke theory (BD) \cite{Brans:1961sx}.
In spite of its simplicity, BD should be a good representative.
This is because the more complicated scalar-tensor theories, such as Horndeski gravity \cite{Horndeski:1974wa}, Degenerate Higher-Order Scalar-Tensor theories \cite{Langlois:2015cwa,Langlois:2015skt,BenAchour:2016fzp} and spatially covariant gravity \cite{Gao:2014soa,Gao:2014fra,Gao:2018znj}, contain more general interactions for the metric field and the scalar field $\varphi$.
None of the interactions should have impact on the asymptotic symmetry or the degenerate vacua.
Therefore, BD should be a good example to study the memory effect in scalar-tensor theories.

As discussed in several previous works, there are three degrees of freedom in the scalar-tensor theory.
Two of them excite the familiar $+$ and $\times$ polarizations as in GR, and the third one, $\varphi$, excites the transverse breathing polarization, if it is massless \cite{Will:2014kxa,Liang:2017ahj,Hou:2017bqj,Gong:2018ybk}. 
Thus, one may call the GW corresponding to the $+$ and $\times$ polarizations the tensor wave, and the one to the breathing polarization the scalar wave.
There are thus two sectors in the scalar-tensor theory: the tensor and the scalar sectors.
One expects that there should be memory effects in both sectors. 

In this work, we would like to find out gravitational memory effects in these two sectors by solving the equations of motion of BD using a generalized Bondi-Sachs coordinate system $\{u,r,\theta,\phi\}$ \cite{Fletcher2003bs}, with $r$ the luminosity radius asymptotically ($r\rightarrow+\infty$).
The resultant metric components and the scalar field $\varphi$ will be expanded in powers of $1/r$.
The BMS symmetry is then determined by looking for the coordinate transformations that preserve the asymptotic behaviors of the metric components and $\varphi$.
It turns out that the BMS transformations in BD are also parameterized by the arbitrary functions $\alpha$ on  a unit 2-sphere, and the smooth conformal vector fields $Y^A$ of the round 2-metric with $A=\theta,\phi$.
Therefore, analogously to GR, the BMS transformations in BD constitute a group, which is also a semi-direct product of the supertranslations (parameterized by $\alpha$) and the Lorentz group (parameterized by $Y^A$).
Then, once the vacua in the tensor and the scalar sectors are properly defined, one discovers that the vacua in both sectors are degenerate. 
Different vacua in the tensor sector are related to each other by supertranslations, as in GR, while those in the scalar sector by Lorentz transformations.
Therefore, similarly to GR, the transition among the tensor vacua is induced by the null energy fluxes through $\scri$, which include the energy flux of the scalar GW if it exists.
This results in the tensor memory effect like the one in GR.
However, the transition among the scalar vacua is induced by the angular momentum fluxes through $\scri$.
As a consequence, the scalar memory effect caused by this transition is absent in GR. 
The charges and the fluxes associated with the BMS symmetries are computed in ref.~\cite{Hou:2020wbo} with Wald-Zoupas formalism \cite{Wald:1999wa} based on Penrose's conformal completion  \cite{Penrose:1962ij,Penrose:1965am}, as opposed to the coordinate method used in ref.~\cite{Tahura:2020vsa}.

The memory effect mentioned above is also specifically called the displacement memory effect, as opposed to the spin memory and the center-of-mass (CM) memory, which are relatively  new in GR \cite{Pasterski:2015tva,Nichols:2018qac}.
The former new effect causes the time delay between two counter-orbiting massless particles in a circular orbit, and might be detected by LISA.
And the later one is related to the changes in the CM part of the angular momentum, which turns out to be difficult to be detected in the near future.
From the results in section~\ref{sec-asys}, one knows that there are also the two new memory effects in BD, specifically, in the tensor sector.
Neither of them exists in the scalar sector.
So in this work, we will not elaborate either on spin memory effect or on CM memory effect.
Finally, in the typical discussion of the displacement memory, one assumes the vanishing of the initial and the final relative velocities of the test particles. 
This might not be the case, and some velocity memory effect exists, e.g., for plane GWs in GR \cite{Zhang:2017rno,Zhang:2017geq,Zhang:2018srn}, which should also be present for more general GWs in an asymptotically flat spacetime. 
We will not discuss this memory effect in this work.

In the past, several works investigated the memory effect in scalar-tensor theories. 
Lang calculated the waveform of the GW generated by a compact binary system \cite{Lang:2013fna,Lang:2014osa}. 
The tensor GW was calculated up to the second post-Newtonian (PN) order beyond the quadrupole approximation, while the scalar GW waveform is accurate up to 1.5 PN order.
The memory effect occurs both in the tensor and the scalar sectors. 
Du and Nishizawa proposed the use of the memory effect as a probe to the nature of gravity, taking the scalar-tensor theory as an example \cite{Du:2016hww}, and the scalar memory effect can be used to test the screening BD theory \cite{Koyama:2020vfc}.
The gravitational memory effect in massive gravity, higher derivative and infinite derivative gravity theories was also studied in refs.~\cite{Kilicarslan:2018bia,Kilicarslan:2018yxd,Kilicarslan:2018unm}.
However, none of these works deciphered the relation between the memory effect and the BMS symmetry.

This work is organized in the following way.
First, the equations of motion are solved in the generalized Bondi-Sachs coordinates in section~\ref{sec-bd}. 
For that end, the boundary conditions for the metric components and the scalar field are determined in section~\ref{sec-bdy}.
Then, the solution is solved for in section~\ref{sec-asys}, and this solution is used to calculate the geodesic deviation equation to reveal the GW polarization content of BD, establishing the connection to the previous works \cite{Liang:2017ahj,Hou:2017bqj,Gong:2018ybk} and memory effects, in section~\ref{sec-pols}.
Second, based on the solution, the BMS transformations are determined in section~\ref{sec-bms}, and the transformation properties of various quantities are also discussed there.
The next section~\ref{sec-mem} focuses on the relation between the memory effect and the BMS symmetry.
After that, there is a brief conclusion in section~\ref{sec-con}.
Appendix~\ref{sec-app-mst} relates the memory effects to the soft theorems, where the memory effect is considered in the scattering of stars in section~\ref{sec-app-bm}, and then, the soft theorems are computed in section~\ref{sec-app-st}.
In this work, the abstract index notation is used \cite{Wald:1984rg}, and the speed of light in vacuum is $c=1$.
Most of the calculation was done with the help of \verb+xAct+ \cite{xact}.

\section{Asymptotically flat spacetimes in Brans-Dicke theory}
\label{sec-bd}

In this section, the asymptotically flat spacetime, i.e., an isolated system, in Brans-Dicke theory is studied.
The action of BD is given by \cite{Brans:1961sx},
\begin{equation}
    \label{eq-act-bd}
    S=\frac{1}{16\pi G_0}\int\ud^4 x\sqrt{-g}\left( \varphi R-\frac{\omega}{\varphi}\nabla_a\varphi\nabla^a\varphi \right)+S_m,
\end{equation}
where $\omega$ is a constant, $G_0$ is the bare gravitational constant, and $S_m$ is the matter action. 
This theory has been studied and well tested for a long time \cite{Will:2014kxa}.
One can generalize the action a bit by adding a general potential term $V(\varphi)$ or making $\omega$ a function of $\varphi$.
Solar-system tests put strong constraints on $\omega\ge4\times10^4$ \cite{Hou:2017cjy}, and the mass of the BD scalar field should be small $10^{-21}<m_s<10^{-15}$ eV \cite{Alsing:2011er}.
BD is also used to explain the inflation \cite{Barrow:1990nv,Clifton:2011jh}, and the current accelerating expansion of the universe \cite{Baccigalupi:2000je,Riazuelo:2001mg,Brax:2004qh}. 
$f(R)$ gravity is dynamically equivalent to BD with a vanishing $\omega$ and a generic potential $V$ \cite{OHanlon:1972xqa, Teyssandier:1983zz}, and is reviewed in ref.~\cite{Sotiriou:2008rp}.
In this work, the simplest BD theory \eqref{eq-act-bd} is considered without any restriction on $\omega$.

The variational principle leads to the following equations of motion,
\begin{subequations}
 \begin{gather}
    R_{ab}-\frac{1}{2}g_{ab}R=\frac{8\pi G_0}{\varphi}(T_{ab}+\mathcal T_{ab}),\label{eq-ein}\\ 
    \nabla_c\nabla^c\varphi=\frac{8\pi G_0}{2\omega+3}T,\label{eq-s}\\ 
    \nabla^b T_{ab}=0,\label{eq-divt}
\end{gather}   
\end{subequations}
in which $T_{ab}=-\frac{2}{\sqrt{-g}}\frac{\delta S_m}{\delta g^{ab}}$ is the matter stress-energy tensor, $T=g^{ab}T_{ab}$ is its trace, and $\mathcal T_{ab}$ is the effective stress-energy tensor for $\varphi$, given by 
\begin{equation}
    \label{eq-eff-s}
    \mathcal T_{ab}=\frac{1}{8\pi G_0}\left[\frac{\omega}{\varphi}\left(\nabla_a\varphi\nabla_b\varphi-\frac{1}{2}g_{ab}\nabla_c\varphi\nabla^c\varphi\right)+\nabla_a\nabla_b\varphi-g_{ab}\nabla_c\nabla^c\varphi\right].
\end{equation}
In fact, eq.~\eqref{eq-divt} follows from eqs.~\eqref{eq-ein} and \eqref{eq-s}.

In a suitable set of coordinate system, e.g., the generalized Bondi-Sachs coordinates $(u,r,x^2=\theta,x^3=\phi)$, the line element is  \cite{Barnich:2010eb}
\begin{equation}
    \label{eq-bc}
    \ud s^2=e^{2\beta}\frac{V}{r}\ud u^2-2e^{2\beta}\ud u\ud r+h_{AB}(\ud x^A-U^A\ud u)(\ud x^B-U^B\ud u),
\end{equation}
where $\beta,\,V,\,U^A$, and $h_{AB}$ ($A,B=2,3$) are six metric functions.
After solving the equations of motion, these metric functions and the scalar field $\varphi$  can be expanded in powers of $1/r$.
For that, one also needs know the boundary conditions, i.e., the asymptotic behaviors of $g_{ab}$ and $\varphi$.

\subsection{Boundary conditions}
\label{sec-bdy}

In GR, it is well-known that \cite{Bondi:1962px,Sachs:1962wk,Barnich:2010eb}
\begin{subequations}
    \label{eq-bdy-gr}
    \begin{gather}
    g_{uu}=-1+\order{r^{-1}},\quad g_{ur}=-1+\order{r^{-2}},\quad g_{uA}=\order{1},\label{eq-exp-gr-1}\\
    g_{rr}=g_{rA}=0,\quad h_{AB}=r^2\gamma_{AB}+\order{r},\label{eq-exp-gr-2}
    \end{gather}
\end{subequations}
where  $\gamma_{AB}$ is the round metric on a unit 2 sphere, 
\begin{equation}
    \label{eq-gmet}
    \gamma_{AB}\ud x^A\ud x^B=\ud\theta^2+\sin^2\theta\ud\phi^2.
\end{equation}
In addition, the determinant of $h_{AB}$ is 
\begin{equation}
    \label{eq-det-gr}
    \det(h_{AB})=r^4\sin^2\theta.
\end{equation}
By this, $r$ is the luminosity radius \cite{Bondi:1962px,Sachs:1962wk}.
In terms of the metric functions, one knows that \cite{Barnich:2010eb}
\begin{equation}
    \label{eq-exp-m}
    \beta=\order{r^{-2}},\quad V=-r+\order{r^0},\quad U^A=\order{r^{-2}}.
\end{equation}
However, in BD, the asymptotic behavior of $g_{ab}$ might be different. 
In fact, one can assume that $\varphi=\varphi_0+\order{r^{-1}}$ with $\varphi_0$ a constant \cite{Alsing:2011er,Hohmann:2015kra,Hou:2017cjy}.
Then, the last two terms in eq.~\eqref{eq-eff-s} begin at the order of $\order{r^{-1}}$, while in GR, the matter stress-energy tensor $T_{ab}$ begins at one order higher, as $T_{ab}$ is usually quadratic in fields. 
Therefore, the asymptotic behavior of $g_{ab}$ in BD should be reanalyzed.

One can perform a conformal transformation $\tilde g_{ab}=\frac{\varphi}{\varphi_0} g_{ab}$ and define $\varphi/\varphi_0=e^{\tilde\varphi}$ such that 
\begin{subequations}
\begin{gather}
    \tilde R_{ab}-\frac{1}{2}\tilde g_{ab}\tilde R=\frac{2\omega+3}{2}\left( \tilde\nabla_a \tilde\varphi\tilde\nabla_b\tilde\varphi-\frac{1}{2}\tilde g_{ab}\tilde\nabla^c\tilde\varphi\tilde\nabla_c\tilde\varphi\right)+8\pi G_0e^{-\tilde\varphi}T_{ab},\label{eq-ein-e}\\ 
    \tilde\nabla_c\tilde\nabla^c\tilde\varphi=\frac{8\pi G_0}{2\omega+3}e^{-\tilde\varphi}\tilde T,\\ 
    \tilde\nabla^b T_{ab}-T_{ab}\tilde\nabla^b\tilde\varphi+\frac{1}{2}\tilde T\tilde\nabla_a\tilde\varphi=0,
\end{gather}
\end{subequations}
where $\tilde T=\tilde g^{ab}T_{ab}$.
Equation~\eqref{eq-ein-e} takes the similar form as in GR with a canonical scalar field $\sqrt{2\omega+3}\tilde\varphi$.
If $\tilde\varphi=\tilde\varphi_0+\order{r^{-1}}$, again with a constant $\tilde\varphi_0$, one expects that $\tilde g_{ab}$ would have the same behavior as the one in GR. 
Therefore, one postulates that in BD, 
\begin{subequations}
    \label{eq-exp-bd}
    \begin{gather}
        g_{uu}=-1+\order{r^{-1}},\quad g_{ur}=-1+\order{r^{-1}},\quad g_{uA}=\order{1},\label{eq-exp-bd-1}\\
        g_{rr}=g_{rA}=0,\quad h_{AB}=r^2\gamma_{AB}+\order{r},\label{eq-exp-bd-2}
    \end{gather}
\end{subequations}
and the determinant condition becomes
\begin{equation}
    \label{eq-det}
    \det(h_{AB})=r^4\left(  \frac{\varphi_0}{\varphi}\right)^2\sin^2\theta.
\end{equation}
So $r$ is no longer the luminosity radius, but it approaches to it near $\scri$.
Comparing eqs.~\eqref{eq-exp-gr-1} with \eqref{eq-exp-bd-1}, one finds out that the condition on $g_{ur}$ is relaxed, so in BD, $\beta=\order{r^{-1}}$.
The remaining metric functions have exactly the same asymptotic behaviors as those in GR, except that the determinant conditions \eqref{eq-det-gr} and \eqref{eq-det} are different.

\subsection{Asymptotic solutions}
\label{sec-asys}

In order to solve the equations of motion near the null infinity, one can first assume 
\begin{subequations}
    \label{eq-exp-in}
 \begin{gather}
    \varphi=\varphi_0+\frac{\varphi_1}{r}+\frac{\varphi_2}{r^2}+\order{\frac{1}{r^3}},\label{eq-exp-s}\\
    h_{AB}=r^2\gamma_{AB}+rc_{AB}+d_{AB}+\order{\frac{1}{r}}.\label{eq-exp-h}
\end{gather}   
\end{subequations}
Here, all of the expansion coefficients are functions of $(u,x^A)$.
The indices of $c_{AB}$ and $d_{AB}$ are lowered and raised by $\gamma_{AB}$ and its inverse $\gamma^{AB}$.
One can check that, the determinant condition \eqref{eq-det} implies that
\begin{subequations}
\begin{gather}
    c_{AB}=\hat c_{AB}-\gamma_{AB}\frac{\varphi_1}{\varphi_0},\\ 
    d_{AB}=\hat d_{AB}+\gamma_{AB}\left(\frac{1}{4}\hat c_{C}^{D}\hat c^{C}_{D}+\frac{\varphi_1^2}{\varphi_0^2}-\frac{\varphi_2}{\varphi_0}\right),
\end{gather}
\end{subequations}
where $\gamma^{AB}\hat c_{AB}=\gamma^{AB}\hat d_{AB}=0$.
So $\hat c_{AB}$ is transverse and traceless, and contains 2 degrees of freedom. 
This is nothing but the transverse-traceless part of the metric perturbation above the flat spacetime background \cite{Liang:2017ahj,Hou:2017bqj,Gong:2018ybk}.
The remaining degree of freedom is encoded in $\varphi_1$.
In the end, the components of the matter stress-energy tensor are assumed to have the following forms,
\begin{subequations}
 \begin{gather}
    T_{uu}=\frac{\bar T_{uu}}{r^2}+\order{\frac{1}{r^3}},\quad 
    T_{ur}=\frac{\bar T_{ur}}{r^4}+\order{\frac{1}{r^5}},\quad 
    T_{rr}=\frac{\bar T_{rr}}{r^4}+\frac{\tilde T_{rr}}{r^5}+\order{\frac{1}{r^6}},\\ 
    T_{uA}=\frac{\bar T_{uA}}{r^2}+\order{\frac{1}{r^3}},\quad 
    T_{rA}=\frac{\bar T_{rA}}{r^3}+\order{\frac{1}{r^4}},\\ 
    T_{AB}=\frac{\bar T}{r}\gamma_{AB}+\frac{\tilde T_{AB}}{r^2}+\order{\frac{1}{r^3}},
\end{gather}   
\end{subequations}
as suggested in ref.~\cite{Flanagan:2015pxa}, based on the behavior of the radiative canonical scalar field in the flat spacetime.
Here, again, all of the expansion coefficients are functions of $(u,x^A)$.
In the following, it is assumed that the matter stress-energy tensor is given.

Following the method presented in ref.~\cite{Barnich:2010eb}, one defines the following quantities,
\begin{subequations}
    \begin{gather}
        k_{AB}=\frac{1}{2}\pd_rh_{AB},\quad k^A_B=h^{AC}k_{BC}=\frac{\delta^A_B}{r}+\frac{\hat K^A_B}{r^2}-\delta^A_B\frac{\pd_r\varphi}{2\varphi},\\ 
        l_{AB}=\frac{1}{2}\pd_uh_{AB},\quad l^A_B=h^{AC}l_{BC}=\frac{\hat L^A_B}{r}-\delta^A_B\frac{\pd_u\varphi}{2\varphi},\\ 
        n_A=\frac{1}{2}e^{-2\beta}h_{AB}\pd_rU^B,\label{eq-def-n}
    \end{gather}
\end{subequations}
where $\hat K^A_A=\hat L^A_A=0$.
$k_{AB}$ is the extrinsic curvature tensor of the null hypersurface $u=\text{const.}$.
In terms of these quantities and the metric functions, the Christoffel symbols can be expressed, but are very complicated, which can be found in ref.~\cite{Barnich:2010eb}.

The equations of motion can be solved in a special order \cite{Bondi:1962px,Sachs:1962wk,Barnich:2010eb,Madler:2016xju}. 
First, one starts with the Einstein's equation \eqref{eq-ein}. 
The simplest one is the $rr$-component, which is
\begin{equation}
    \label{eq-ein-rr}
    \frac{4\pd_r\beta}{r}-\frac{\hat K^A_B\hat K_A^B}{r^4}-\frac{2\pd_r\varphi}{r\varphi}-\frac{2\omega+3}{2}\left( \frac{\pd_r\varphi}{\varphi^2} \right)^2=\frac{8\pi G_0}{\varphi}T_{rr}.
\end{equation}
From this, one can solve for $\pd_r\beta$, and further integrate the resulting equation to give rise to $\beta$, which has the following form,
\begin{equation}
    \label{eq-beta-ser}
    \beta=\frac{\beta_1}{r}+\frac{\beta_2}{r^2}+\order{\frac{1}{r^3}},
\end{equation}
with 
\begin{subequations}
    \begin{gather}
    \beta_1=-\frac{\varphi_1}{2\varphi_0},\\
    \beta_2=-\frac{\hat c_{A}^{B}\hat c^{A}_{B}}{32}-\frac{\pi G_0\bar T_{rr}}{\varphi_0}+\frac{1-2\omega}{16}\left( \frac{\varphi_1}{\varphi_0} \right)^2-\frac{\varphi_2}{2\varphi_0}. 
    \end{gather}
\end{subequations}
The $rA$-component of eq.~\eqref{eq-ein} is also very simple, which can be rearranged to have the following form
\begin{equation}
    \label{eq-ein-rA}
    \begin{split}
    \pd_r(r^2n_A)=&r^2\mathcal D_A\pd_r\beta-2r\mathcal D_A\beta-\mathcal D_B\hat K^B_A+\frac{8\pi G_0}{\varphi}r^2T_{rA}\\ 
    &-r\frac{\mathcal D_A\varphi}{\varphi}-\frac{\hat K^B_A\mathcal D_B\varphi}{\varphi}+(1+\omega)r^2\frac{(\pd_r\varphi)\mathcal D_A\varphi}{\varphi^2},
    \end{split}
\end{equation}
where $\mathcal D_A$ is the compatible covariant derivative for $h_{AB}$.
Integrating this equation leads to the series expansion for $n_A$,
\begin{equation}
    \label{eq-n-series}
    n_A=\frac{\mathscr D_B\hat c^B_A}{2r}+\frac{1}{r^2}\left( N_A+\mathscr U_A\ln r\right)
    +\order{\frac{1}{r^3}},
\end{equation}
where $N_A$ is an integration function and can be named the Bondi angular momentum aspect, and 
\begin{equation}
    \label{eq-def-u}
    \mathscr U_A=\mathscr D_B\hat d_A^B+\frac{4\pi G_0}{\varphi_0}(2\bar T_{rA}+\mathscr D_A\bar T_{rr})+\mathscr D_B\left(\frac{\varphi_1}{\varphi_0}\hat c^B_A  \right),
\end{equation}
with $\mathscr D_A$ the covariant derivative of $\gamma_{AB}$.
By the definition \eqref{eq-def-n}, one can obtain the series expansion for $U^A$, i.e., 
\begin{equation}
    \label{eq-U-series}
    \begin{split}
    U^A=&-\frac{\mathscr D_BC^{AB}}{2r^2}+\frac{1}{r^3}\left[ -\frac{2}{3}N^A+\frac{1}{3}\hat c^{AB}\mathscr D_C\hat c_B^C-\frac{2}{3}\left( \frac{1}{3}+\ln r \right)\mathscr U^A \right]+\order{\frac{1}{r^4}},
    \end{split}
\end{equation}
where $\mathscr U^A=\gamma^{AB}\mathscr U_B$.
Now, it is ready to use the $ur$-component of eq.~\eqref{eq-ein}, whose explicit form is very complicated. 
However, one can combine it with eqs.~\eqref{eq-ein-rr} and \eqref{eq-ein-rA} in a suitable way to obtain a fairly manageable expression,
\begin{equation}
    \label{eq-ein-ur}
    \begin{split}
        \pd_rV-\frac{4\pi G_0}{\varphi}rT_{rr}V=&J\\ 
        =&-2r(\mathcal D_AU^A+ U^A\mathcal D_A\ln\varphi)+\frac{8\pi G_0}{\varphi}r^2(T_{ur}+T_{rA}U^A)\\ 
        &+r^2e^{2\beta}\left[ -\frac{\mathcal R}{2}-\mathcal D_An^A+(n^A+\mathcal D^A\beta)(n_A+\mathcal D_A\beta)+\mathcal D_A\mathcal D^A\beta \right. \\ 
        &\left.-n^A\mathcal D_A\ln\varphi+\mathcal D^A\beta\mathcal D_A\ln\varphi+\frac{1}{2}\mathcal D^A\ln\varphi\mathcal D_A\ln\varphi+\frac{\mathcal D_A\mathcal D^A\varphi}{\varphi}\right]\\ 
        =&-1+\order{r^{-2}},
    \end{split}
\end{equation}
where $\mathcal R$ is the Ricci scalar for $h_{AB}$, and the series expansions for $\beta,\,\phi,\,U^A,\,n_A$, and $h_{AB}$ have been submitted to obtain the last line.
With this, $V$ can be solved for, that is, 
\begin{equation}
    \label{eq-V-sol}
    \begin{split}
        V=2m+\int^r_\infty \left( J+\frac{4\pi G_0}{\varphi}rT_{rr}V \right)_{r'}\ud r',
    \end{split}
\end{equation}
where $m$ is an integration function, called the Bondi mass aspect \cite{Madler:2016xju}, and the subscript $r'$ of the round brackets means to evaluate the integrand at $r'$.
The form of this equation suggests that one can replace $V$ in the integration by the left-hand side, and repeat for enough times to obtain a series expansion for $V$ truncated at a desired order.
Thus, one finds out that 
\begin{equation}
    \label{eq-V-series}
    V=-r+2m+\order{r^{-1}}.
\end{equation}
Next, one considers the $AB$-component of eq.~\eqref{eq-ein}, which is even more complicated. 
After some tedious manipulation, one arrives at 
\begin{equation}
    \label{eq-ein-AB}
    \begin{split}
    \pd_r\hat L^A_B+\mathcal H^{AD}_{BC}\hat L^C_D=&J^A_B\\ 
        =&r\left[e^{2\beta}\left( n^An_B+\mathcal D^A\beta\mathcal D_B\beta+\mathcal D^A\mathcal D_B\beta-\frac{\mathcal R^A_B}{2}+\frac{4\pi G_0}{\varphi}T^A_B\right.\right.\\ 
        &\left.\left.+\frac{\omega}{2\varphi^2}\mathcal D^A\varphi\mathcal D_B\varphi+\frac{\mathcal D^A\mathcal D_B\varphi}{2\varphi} \right)-\frac{1}{2}\mathcal D^A(e^{2\beta}n_B)-\frac{1}{2}\mathcal D_B(e^{2\beta}n^A)\right]\\ 
        &-\frac{1}{2}(\mathcal D^AU_B+\mathcal D_BU^A)+\frac{1}{r}\left[ \hat K^{AC}\mathcal D_{[C}U_{B]}+\hat K_{BC}\mathcal D^{[C}U^{A]}\right.\\ 
        &\left.-\frac{1}{2}\hat K^A_B\mathcal D_CU^C-U^C\mathcal D_C\hat K^A_B -\frac{1}{2\varphi}\hat K^A_BU^C\mathcal D_C\varphi+\pd_r\left(\frac{V}{2r}\hat K^A_B \right)\right]\\ 
        &+\delta^A_B\left\{r\left[ \frac{2\pi G_0}{\varphi}(2T_{ur}+2U^CT_{rC}-e^{2\beta} T^C_C)-e^{2\beta}\frac{n^C\mathcal D_C\varphi}{2\varphi}\right.\right.\\
        &\left.\left. +\frac{\mathcal D_C(e^{2\beta}\mathcal D^C\varphi)}{4\varphi}\right]-\frac{\mathcal D_CU^C}{2}-\frac{U^C\mathcal D_C\varphi}{\varphi}+\frac{2\pi G_0}{\varphi}VT_{rr}-\frac{\pd_rV}{2r}\right\},
    \end{split}
\end{equation}
where $\mathcal R^A_B$ is the Ricci tensor for $h_{AB}$ with one index raised, and 
\begin{equation}
    \label{eq-def-o}
    \mathcal H^{AD}_{BC}=\frac{1}{r^2}(\hat K^A_C\delta^D_B-\hat K^D_B\delta^A_C).
\end{equation}
The both sides of eq.~\eqref{eq-ein-AB} can be rewritten using the series expansions of the metric functions and $\varphi$, i.e., 
\begin{equation}
    \label{eq-ein-AB-exp}
    \begin{split}
    -\frac{\pd_u(\hat d^A_B+\varphi_1\hat c^B_A/\varphi_0)}{2r^2}+\order{\frac{1}{r^3}}=&\frac{1}{4r^2}(2\hat c_B^A-\mathscr D_C\mathscr D_B\hat c^{AC}-\mathscr D^C\mathscr D^A\hat c_{BC}+\mathscr D^2\hat c^A_B\\ 
        &+\delta^A_B\mathscr D_C\mathscr D_D\hat c^{CD})+\order{\frac{1}{r^3}}.
    \end{split}
\end{equation}
It turns out that the terms in the brackets on the right-hand side add up to zero \cite{Barnich:2010eb}, so 
\begin{equation}
    \label{eq-dud}
    \pd_u\hat d_{AB}=-\pd_u\left( \frac{\varphi_1}{\varphi_0}\hat c_{AB} \right).
\end{equation}
One can also try to integrate eq.~\eqref{eq-ein-AB}, 
\begin{equation}
    \label{eq-L-sol}
    \hat L^A_B=-\frac{N^A_B}{2}+\int_\infty^r\left( J^A_B-\mathcal H^{AD}_{BC}\hat L^C_D \right)_{r'}\ud r',
\end{equation}
with $N^A_B$ a new integration function,
and replace $\hat L^C_D$ in the brackets with the right-hand side of the equation several times in order to obtain the series expansion for $\hat L^A_B$ up to an appropriate order. 
It turns out that, this treatment leads to an simple result, that is,
\begin{equation}
    \label{eq-def-news}
    N_{AB}=\gamma_{AC}N^C_B=-\pd_u\hat c_{AB},
\end{equation}
which is called the news tensor \cite{Geroch1977,Ashtekar:1981hw}.
This is also a symmetric and traceless tensor, i.e., $N_{AB}=N_{BA}$ and $\gamma^{AB}N_{AB}=0$.
So it has two degrees of freedom, like $\hat c_{AB}$.
The nonvanishing of it signals the presence of the tensor GW.

Up to now, the series expansions for all the metric functions have been determined, i.e., \eqref{eq-exp-h}, \eqref{eq-beta-ser}, \eqref{eq-U-series} and \eqref{eq-V-series}. 
It is now ready to make use of the remaining equations. 
First, the $uu$-component of the Einstein's equation \eqref{eq-ein} gives the evolution of the Bondi mass aspect,
\begin{equation}
    \label{eq-uu2-mdot}
    \dot {m}=-\frac{4\pi G_0}{\varphi_0}\bar T_{uu}-\frac{1}{4}\mathscr D_A\mathscr D_BN^{AB}-\frac{1}{8}N_{AB}N^{AB}-\frac{2\omega+3}{4}\left( \frac{N}{\varphi_0} \right)^2,
\end{equation}
where $N=\dot\varphi_1$, and dot means the partial $u$ derivative.
Second, the $uA$-component of eq.~\eqref{eq-ein} gives rise to the evolution of the Bondi angular momentum aspect,
\begin{equation}
    \label{eq-uA2-ndot}
    \begin{split}
    \dot {N}_A=&-\frac{8\pi G_0}{\varphi_0}\bar T_{uA}+\frac{\pi G_0}{\varphi_0}\mathscr D_A\pd_u\bar T_{rr}+\mathscr D_Am\\
        &+\frac{1}{4}(\mathscr D_B\mathscr D_A\mathscr D_C\hat c^{BC}-\mathscr D_B\mathscr D^B\mathscr D_C\hat c_A^C)\\ 
        &-\frac{1}{16}\mathscr D_A(N^{B}_{C}\hat c_{B}^{C})+\frac{1}{4}N^{B}_{C}\mathscr D_A\hat c_{B}^{C}+\frac{1}{4}\mathscr D_B(N_{A}^{C}\hat c^{B}_{C}-\hat c_{A}^{C}N^{B}_{C})\\ 
        &+\frac{2\omega+3}{8\varphi_0^2}(\varphi_1\mathscr D_AN-3N\mathscr D_A\varphi_1).
    \end{split}
\end{equation}
Third, the equation of motion \eqref{eq-s} for  $\varphi$ leads to 
\begin{equation}
    \label{eq-ev-varphis}
    \dot\varphi_2=\frac{\varphi_1N}{\varphi_0}-\frac{1}{2}\mathscr D^2\varphi_1+\frac{8\pi G_0}{2\omega+3}\bar T.
\end{equation}
Similarly to $N_{AB}$, a nonvanishing $N$ implies that there is the scalar GW at $\scri$.
Finally, eq.~\eqref{eq-divt} results in
\begin{equation}
    \label{eq-mst-ev}
    \pd_u\bar T_{rA}=\mathscr D_A\bar T,\quad
    \pd_u\bar T_{rr}=-2\bar T.
\end{equation}
The equations giving rise to the above results are extremely complicated and not very illuminating, so none of them is displayed. 

Reflecting upon the computation, one realizes that the news tensor $N_{AB}$ and the scalar $N=\dot\varphi_1$ play special roles. 
As long as $N_{AB}(u,x^C)$ and $N(u,x^A)$ are prescribed on $\scri$, one can determine $\hat c_{AB}(u,x^C)$ and $\varphi_1(u,x^A)$, given their initial values $\hat c_{AB}(u_0,x^C)$ and $\varphi_1(u_0,x^A)$ at a certain retarded time $u_0$.
Then by eqs.~\eqref{eq-uu2-mdot} and \eqref{eq-uA2-ndot}, the evolutions of the mass and the angular momentum aspects, $m(u,x^A)$ and $N_A(u,x^B)$, are known with the knowledge of their initial values $m(u_0,x^A)$ and $N_A(u_0,x^B)$, respectively.
Similarly, eq.~\eqref{eq-dud} also gives the evolution of $\hat d_{AB}(u,x^C)$, again once its initial value $\hat d_{AB}(u_0,x^C)$ is given.
In the end, eq.~\eqref{eq-ev-varphis} can now be used to fix $\varphi_2(u,x^A)$  with $\varphi_2(u_0,x^A)$  provided.
Indeed, by their definitions, $N_{AB}$ and $N$ are directly associated with the degrees of freedom in BD.

Therefore, the metric components are completely determined up to certain orders in $1/r$,
\begin{subequations}
    \label{eq-met-sol}
    \begin{gather}
        g_{uu}=-1+\frac{2m+\varphi_1/\varphi_0}{r}+\order{\frac{1}{r^2}},\\
     g_{ur}=-1+\frac{\varphi_1}{\varphi_0r}+\frac{1}{r^2}\left[ \frac{1}{16}\hat c_{A}^{B}\hat c^{A}_{B}+\frac{2\omega-5}{8}\left( \frac{\varphi_1}{\varphi_0} \right)^2+\frac{\varphi_2}{\varphi_0}+\frac{2\pi G_0}{\varphi_0}\bar T_{rr} \right]+\order{\frac{1}{r^3}},\\ 
     \begin{split}
         g_{uA}=&\frac{\mathscr D_B\hat c^B_A}{2}+\frac{2}{3r}\left[ N_A+\frac{1}{4}\hat c_{AB}\mathscr D_C\hat c^{BC}+\left( \frac{1}{3}+\ln r \right)\mathscr U_A-\frac{\varphi_1}{12\varphi_0}\mathscr D_B\hat c^B_A \right]\\ 
         &+\order{\frac{1}{r^2}},
     \end{split}\label{eq-met-uA}\\
     \begin{split}
     g_{AB}=&r^2\gamma_{AB}+r\left( \hat c_{AB}-\gamma_{AB}\frac{\varphi_1}{\varphi_0} \right)+\hat d_{AB}+\gamma_{AB}\left( \frac{1}{4}\hat c_{C}^{D}\hat c^{C}_{D}+\frac{\varphi_1^2}{\varphi_0^2}-\frac{\varphi_2}{\varphi_0} \right)\\
     &+\order{\frac{1}{r}}.
     \end{split}
    \end{gather}
\end{subequations}
Of course, $\varphi$ has been expanded in eq.~\eqref{eq-exp-s} with the expansion coefficients satisfying eq.~\eqref{eq-ev-varphis}.
Note the presence of the term with $\ln r$ in $g_{uA}$.
This leads to the existence of singularities on the unit 2-sphere \cite{Barnich:2010eb}. 
To avoid that, we would like to set $\mathscr U_A=0$, so 
\begin{equation}
   \label{eq-noln} 
    \mathscr D_B\hat d_A^B=-\mathscr D_B\left( \frac{\varphi_1}{\varphi_0}\hat c_A^B \right)-\frac{4\pi G_0}{\varphi_0}(2\bar T_{rA}+\mathscr D_A\bar T_{rr}),
\end{equation}
which is consistent with eqs.~\eqref{eq-dud} and \eqref{eq-mst-ev}.

When the scalar profile is trivial, i.e., $\varphi_n=0$ for $n\ge1$, the metric components \eqref{eq-met-sol} reduce to the GR's results \cite{Barnich:2010eb}, as one would expect.
Now, a close inspection reveals that the coefficient of the $1/r$ term of $g_{uu}$ is  $m'=m+\varphi_1/2\varphi_0$.
Here, we do not call $m'$ the Bondi mass aspect as in GR, because the evolution equation \eqref{eq-uu2-mdot} suggests that $m$ is consistent the original meaning of the Bondi mass aspect. 
Indeed, it is well-known that the outgoing GWs carry energy away from an isolated system, so a certain ``mass'' of the system should decrease.
Now, define the following masses
\begin{subequations}
    \begin{gather}
        M(u)=\frac{\varphi_0}{4\pi G_0}\oint m(u,x^A)\sin\theta\ud\theta\ud\phi,\\ 
        M'(u)=\frac{\varphi_0}{4\pi G_0}\oint m'(u,x^A)\sin\theta\ud\theta\ud\phi,
    \end{gather}
\end{subequations}
where the integrations are over the unit 2-sphere.
Then, $M$ decays according to
    \begin{equation}
        \dot M=-\oint \left( \bar T_{uu}+\frac{\varphi_0}{32\pi G_0}N_{AB}N^{AB}+\frac{2\omega+3}{16\pi G_0\varphi_0}N^2 \right)\sin\theta\ud\theta\ud\phi,
    \end{equation}
in which the second term in eq.~\eqref{eq-uu2-mdot} drops out because it is a total divergence.
The first term in the square brackets above is the energy flux of the matter fields, such as the electromagnetic field. 
The second one represents the energy flux of the tensor GW, and the third is for the scalar GW \cite{Alsing:2011er,Hou:2017cjy}.
However, the rate of the decrease in $M'(u)$ is 
\begin{equation}
    \dot M'=\dot M+\frac{1}{8\pi G_0}\oint N\sin\theta\ud\theta\ud\phi,
\end{equation}
so the integrand in the second term is linear in $N$, which is not the energy flux of any sort. 
Therefore, this justifies that $m$ should be called the Bondi mass aspect.
Moreover, the discussion in ref.~\cite{Hou:2020wbo} also suggests that $M(u)$ is \textit{the} Bondi mass.
In fact, $M'$ is the Arnowitt-Deser-Misner mass \cite{Arnowitt:1962hi}, which can be understood by performing the coordinate transformation $u=t-r+\order{r^{-1}}$.

As a final remark, it is known that the Kaluza-Klein reduction of a higher-dimensional Einstein gravity results in scalar-tensor theories \cite{Overduin:1998pn}. 
Since the higher-dimensional asymptotically flat spacetimes have been solved for in the past \cite{Tanabe:2009va,Tanabe:2011es,Hollands:2013cva,Hollands:2016oma}, one may expect that the solutions presented here can be recovered by the reduction. 
However, we will not perform the reduction in this work, because of the complication. 
According to the reduction procedure \cite{Overduin:1998pn}, one would like to assume that the higher-dimensional spacetime metric possesses a number of exact symmetries.
This puts some constraints on the higher-dimensional asymptotically flat spacetime, and it is not trivial to express the constraints in terms of the Bondi-Sachs coordinates, which makes the reduction mathematically very involved.
In addition, the reduction not only gives rise to some scalar fields, but also some vector fields, and they couple to each other in some nontrivial ways \cite{Overduin:1998pn}. 
However, what we have considered is a theory in which the BD scalar field only couples to the metric, then one should be careful in interpreting the results of the reduction.
Because of these complications, we would consider the Kaluza-Klein reduction in the future work.

\subsection{Gravitational wave polarizations}
\label{sec-pols}

In the previous sections, the asymptotic solution of an isolated system in BD has been calculated. 
It was claimed that when $N_{AB}$ and $N$ are nonvanishing, there exist GWs at $\scri$.
Here, one can verify this statement.
Previously, several works discussed GWs and the polarizations in scalar-tensor theories \cite{Liang:2017ahj,Hou:2017bqj,Gong:2017bru,Gong:2018ybk,Hou:2018mey}.
It is claimed that in a scalar-tensor theory with a massless scalar field, there are three polarizations. 
Two of them are the familiar $+$ and $\times$ polarizations, and the third one is the transverse breathing polarization. 
Now, let us examine what they are in terms of $N_{AB}$ and $N$, or $\hat c_{AB}$ and $\varphi_1$.
In fact, by simply counting the degrees of freedom, one immediately realizes that $\hat c_{AB}$ excites the $+$ and $\times$ polarizations, while $\varphi_1$ the transverse breathing one.

The GW is detected based on the geodesic deviation equation \cite{Wald:1984rg},
\begin{equation}
    \label{eq-gde}
    T^c\nabla_c(T^b\nabla_bS^a)=-R_{cbd}{}^aT^cS^bT^d,
\end{equation}
where $T^a$ is the 4-velocity of a freely falling test particle, and $S^a$ is the deviation vector between adjacent test particles.
To detect the GW of an isolated system, one places test particles far away from the isolated system. 
There are many ways to choose test particles. 
One of them is to place them at a fixed radius $r_0$ and the fixed direction $x^A_0$, and these particles are called BMS detectors \cite{Strominger:2014pwa}.
In general, these particles will be accelerated, but as long as $r_0$ is very large, they approximately freely fall. 
In terms of the coordinates, one has \cite{Strominger:2014pwa}
\begin{equation}
    X^{u,r}_\text{BMS}(s)=X^{u,r}_\text{geo}(s)+\order{r_0^{-1}},\quad x^A_\text{BMS}(s)=x^A_\text{geo}(s)+\order{r_0^{-2}},
\end{equation} 
where the subscript ``BMS'' means the coordinates are for BMS detectors, and ``geo'' for the freely falling particles.
So the 4-velocity of the BMS detector is approximately $T^a=(\pd_u)^a$ at a far distance $r_0$.
Define $(e_{\hat \theta})^a=r^{-1}(\pd/\pd \theta)^a$, and $(e_{\hat\phi})^a=(r\sin\theta)^{-1}(\pd/\pd\phi)^a$. 
One can check that they are unit spatial vectors, normal to each other and to $T^a$, at $\scri$.
Then, let $S^a=S^{\hat A}(e_{\hat A})^a$, and the geodesic deviation equation is approximately,
\begin{equation}
    \label{eq-gde-c}
    \ddot S^{\hat A}\approx-R_{u\hat Bu}{}^{\hat A}S^{\hat B},
\end{equation}
where the electric part of the Riemann tensor $R_{abc}{}^d$ is 
\begin{equation}
    \label{eq-el-rie}
    R_{u\hat Au\hat B}=-\frac{1}{2r}\left( \pd_u^2\hat c_{\hat A\hat B}-\gamma_{\hat A\hat B}\frac{\pd_u^2\varphi_1}{\varphi_0} \right)+\order{\frac{1}{r^2}}.
\end{equation}
Here, $\gamma_{\hat A\hat B}=\delta_{\hat A\hat B}$.
It is instructive to compare eq.~\eqref{eq-el-rie} with the matrix (29) in ref.~\cite{Hou:2017bqj}.
To make a sensible comparison, one must understand that eq.~\eqref{eq-el-rie} gives only the transverse part of the matrix (29), i.e., the top-left $2\times2$ submatrix.
The remaining part of the matrix (29) is related to $R_{u\hat Au\hat z}$ and $R_{u\hat zu\hat z}$ with $\hat z$ direction parallel to $-(\pd_u)^a+(\pd_r)^a$ near $\scri$.
These Riemann tensor components are of higher orders in $1/r$, so one can ignore them for the purpose of comparison.
So one clearly recognizes that it is $\hat c_{AB}$ that is responsible for the presence of the $+$ and $\times$ polarizations, and $\varphi_1$ for the transverse breathing one.
When $\pd_u\hat c_{AB}=0$ and $\dot\varphi_1=0$, that is, $N_{AB}=0$ and $N=0$, there is no change in the deviation, so there are no GWs.

Now, integrating eq.~\eqref{eq-gde-c} twice results in 
\begin{equation}
    \label{eq-dev-c}
    \Delta S_{\hat A}\approx\frac{1}{2r}\left( \Delta\hat c_{\hat A\hat B}-\delta_{\hat A\hat B}\frac{\Delta\varphi_1}{\varphi_0} \right)S_{0}^{\hat B}+\order{\frac{1}{r^2}},
\end{equation}
in which $S^{\hat B}_0$ is the initial deviation vector at the retarded time $u_0$ when there were no GWs, i.e., $N_{AB}(u_0,x^C)=0$ and $N(u_0,x^A)=0$.
A radiating isolated system will eventually settle down to a state in which no GWs can be emitted, and so $N_{AB}$ and $N$ vanish again. 
But the deviation vector $S^{\hat A}$ may not return to its initial value, that is, 
\begin{equation}
    \label{eq-def-mm}
    \Delta S_{\hat A}\ne0.
\end{equation}
This means that there might exist a permanent change in the relative distances between test particles.
This is the GW memory effect.
The memory associated with $\Delta\hat c_{\hat A\hat B}$ would be similar to the one in GR, as shown below. 
The one associated with $\Delta\varphi_1$ is new, and named S memory in ref.~\cite{Du:2016hww}.
Since the scalar GW causes a uniform expansion and contraction in the plane perpendicular to the propagation direction, S memory will also be uniform in that plane. 
Because the interferometer responses to the GWs depending on the propagation direction and the polarization, it is possible to detect the tensor and scalar polarizations, and thus S memory, by combining data from three or more interferometers \cite{Isi:2015cva,Isi:2017fbj,Du:2016hww,Koyama:2020vfc}.

In GR, memory effect is closely related to the asymptotic symmetry that preserves the boundary conditions \eqref{eq-bdy-gr} and the determinant condition \eqref{eq-det-gr}.
In the following, the relation between the memory effect and the asymptotic symmetry in BD is discussed. 
One begins with the asymptotic symmetry in BD.

\section{Asymptotic symmetry}
\label{sec-bms}

Like GR, the scalar-tensor theory is diffeomorphic. 
Among all the coordinate transformations, there exists a set of transformations that preserve the boundary conditions \eqref{eq-exp-bd} and \eqref{eq-exp-s}.
Such a set forms a group, and is called the BMS group \cite{Bondi:1962px,Sachs:1962wk,Sachs:1962zza}.
Let $\xi^\mu$ be a vector field that generates an infinitesimal BMS transformation, then one requires that \cite{Barnich:2010eb}
\begin{subequations}
    \begin{gather}
        \lie_\xi g_{rr}=\lie_\xi g_{rA}=0,\label{eq-bms-inf-0}\\
         g^{AB}\lie_\xi g_{AB}=-\frac{2\lie_\xi\varphi}{\varphi},\label{eq-bms-inf-1}\\
        \lie_\xi g_{ur}=\order{r^{-1}},\quad \lie_\xi g_{uA}=\order{1},\quad \lie_\xi g_{AB}=\order{r},\label{eq-bms-inf-2}\\ 
        \lie_\xi g_{uu}=\order{r^{-1}},\label{eq-bms-inf-3}\\
        \lie_\xi\varphi=\order{r^{-1}}.\label{eq-bms-inf-s}
    \end{gather}
\end{subequations}
In GR, one usually demands that $g^{AB}\lie_\xi g_{AB}=0$, but here, we impose a different condition given by eq.~\eqref{eq-bms-inf-1}.
This is suggested by the determinant condition \eqref{eq-det}, and would reproduce GR's condition when $\varphi$ is trivial.
By some tedious calculation, one finds out that eqs.~\eqref{eq-bms-inf-0} and \eqref{eq-bms-inf-1} determine the components of $\xi^\mu$, i.e.,
\begin{subequations}
    \begin{gather}
        \xi^u=f(u,x^A),\label{eq-xi-u}\\ 
        \xi^A=Y^A(u,x^A)-(\mathscr D_Bf)\int_r^\infty e^{2\beta}g^{AB}\ud r',\label{eq-xi-a}\\
        \xi^r=\frac{r}{2}(U^A\mathscr D_Af-\mathscr D_A\xi^A),\label{eq-xi-r}
    \end{gather}
\end{subequations}
where $f$ and $Y^A$ are arbitrary integration functions.
Next, eq.~\eqref{eq-bms-inf-2} implies the following properties of the integration functions,
\begin{subequations}\label{eq-pro-fy}
    \begin{gather}
        \dot f=\frac{\psi}{2}=\frac{1}{2}\mathscr D_AY^A,\label{eq-pro-f}\\ 
        \dot Y^A=0,\label{eq-pro-y}\\ 
        \lie_Y\gamma_{AB}=\psi\gamma_{AB}.\label{eq-pro-conf}
    \end{gather}
\end{subequations}
Finally, eqs.~\eqref{eq-bms-inf-3} and \eqref{eq-bms-inf-s} lead to no new restrictions on $\xi^\mu$.
Equation~\eqref{eq-pro-y} shows that $Y^A$ depends only on $x^A$, and it is actually a conformal Killing vector field for $\gamma_{AB}$ according to eq.~\eqref{eq-pro-conf}.
By eq.~\eqref{eq-pro-f}, one obtains that 
\begin{equation}
    \label{eq-f-int}
    f=\alpha(x^A)+u\frac{\psi}{2}
\end{equation}
with $\alpha(x^A)$ an integration function of $x^A$.

Formally, eqs.~\eqref{eq-xi-u} and \eqref{eq-xi-a} coincide with those in GR \cite{Barnich:2010eb}, because, in both GR and BD, it is demanded that $g_{rr}=g_{rA}=0$ in any Bondi coordinates. 
Although the two theories have distinct determinant conditions, $\xi^r$ [refer to \eqref{eq-xi-r}] still takes the same form as that in GR due to eq.~\eqref{eq-bms-inf-1}.
The properties \eqref{eq-pro-fy} and \eqref{eq-f-int} satisfied by $f$ and $Y^A$ are also the same as those in GR.
Using the asymptotic expansions of the metric functions and the scalar field, one knows that 
\begin{subequations}\label{eq-exp-xi}
    \begin{gather}
        \xi^A=Y^A-\frac{\mathscr D^Af}{r}+\frac{\hat c^{AB}\mathscr D_Bf}{2r^2}+\order{\frac{1}{r^3}},\\ 
        \xi^r=-\frac{r}{2}\psi+\frac{1}{2}\mathscr D^2f-\frac{1}{2r}\left[ (\mathscr D_Af)\mathscr D_B\hat c^{AB}+\frac{1}{2}\hat c^{AB}\mathscr D_A\mathscr D_Bf \right]+\order{\frac{1}{r^2}}.
    \end{gather}
\end{subequations}
Despite the fact that eqs.~\eqref{eq-exp-xi} contains no expansion coefficients $\varphi_n$ of $\varphi$, and thus takes the same form as those in GR, $\varphi_n$ show up at higher order terms in $1/r$.
These higher order terms are very complicated and not very illuminating, so none is displayed. 

As shown by eqs.~\eqref{eq-exp-xi}, together with eq.~\eqref{eq-xi-u}, an infinitesimal BMS transformation is generated by the arbitrary functions $f$ and $Y^A$, or $\alpha$ and $Y^A$, referring to eq.~\eqref{eq-f-int}.
In particular, the transformation generated by $\alpha\ne0$ and $Y^A=0$ is called a supertranslation.
The supertranslations form a normal subgroup of the BMS group, denoted by $\mathscr S$.
Among the supertranslations, those generated by 
\begin{equation}
    \label{eq-a-trans}
    \alpha=\alpha_0+\alpha_1\sin\theta\cos\phi+\alpha_2\sin\theta\sin\phi+\alpha_3\cos\theta,
\end{equation}
with $\alpha_n\;(n=0,1,2,3)$ constant are translations, and constitute the translation subgroup $\mathscr T$.
The transformations generated by $\alpha=0$ and $Y^A\ne0$ are special. 
Since $Y^A$ is a conformal Killing vector field on a unit 2 sphere, if it is required to be smooth and finite, the set of such vector fields is isomorphic to the Lorentz algebra. 
So in this case, the transformations generated by such conformal Killing vector fields are called the Lorentz transformations, and the corresponding group is labeled by $\mathscr L$.
One may drop the requirement that $Y^A$ be smooth and finite, and the resulting transformations are called the super-rotations or superboosts \cite{Banks:2003vp,Barnich:2009se,Barnich:2011ct,Barnich:2011mi,Kapec:2014opa,Flanagan:2015pxa}. 
In this work, we will not consider the super-rotations/superboosts.
For their associated memory effects and soft theorems in GR, please refers to refs.~\cite{Compere:2018ylh,Ruzziconi:2019pzd}. 

Under an infinitesimal BMS transformation $\xi^a$, the scalar field changes according to $\delta_\xi\varphi=\lie_\xi\varphi$, and so 
\begin{subequations}
    \label{eq-bms-phis}
    \begin{gather}
    \delta_\xi\varphi_1=fN+\frac{\psi}{2}\varphi_1+Y^A\mathscr D_A\varphi_1,    \label{eq-bms-phi1}\\
    \delta_\xi\varphi_2=f\dot\varphi_2-\frac{\mathscr D^2f}{2}\varphi_1-(\mathscr D^Af)\mathscr D_A\varphi_1+\psi\varphi_2+Y^A\mathscr D_A\varphi_2.        
    \end{gather}
\end{subequations}
At the same time, the metric transforms according to $\delta_\xi g_{ab}=\lie_\xi g_{ab}$, from which, we find out that 
\begin{subequations}
    \label{eq-bms-mets}
    \begin{gather}
        \delta_\xi\hat c_{AB}=-fN_{AB}-2\mathscr D_A\mathscr D_Bf+\gamma_{AB}\mathscr D^2f+\lie_Y\hat c_{AB}-\frac{\psi}{2}\hat c_{AB},\label{eq-bms-c}\\
        \delta_\xi\hat d_{AB}=f\pd_u\hat d_{AB}+\lie_Y\hat d_{AB}+\frac{\varphi_1}{\varphi_0}(2\mathscr D_A\mathscr D_Bf-\gamma_{AB}\mathscr D^2f),\\
        \begin{split}
        \delta_\xi m=&f\dot m+Y^A\mathscr D_Am+\frac{3}{2}\psi m-\frac{1}{4}N^{AB}\mathscr D_A\mathscr D_Bf\\ 
        &-\frac{1}{2}(\mathscr D_Af)\mathscr D_BN^{AB}+\frac{1}{8}\hat c^{AB}\mathscr D_A\mathscr D_B\psi+\frac{\varphi_1}{4\varphi_0}(\mathscr D^2\psi+2\psi),
        \end{split}
    \end{gather}
    \begin{equation}
         \begin{split}
        \delta_\xi N_A=&f\dot N_A+\lie_Y N_A+\psi N_A+3m\mathscr D_Af\\
           &-\frac{1}{2}\hat d_A^B\mathscr D_B\psi-\frac{2\pi G_0}{\varphi_0}\bar T\mathscr D_Af-\frac{\pi G_0}{\varphi_0}\bar T_{rr}\mathscr D_A\psi\\ 
           &-\frac{1}{32}\hat c_{C}^{B}\hat c^{C}_{B}\mathscr D_A\psi-\frac{1}{2}\hat c_{A}^{B}N_{B}^{C}\mathscr D_Cf+\frac{3}{16}\hat c_{B}^{C}N^{B}_{C}\mathscr D_Af\\ 
           &+\frac{\hat c_A^B}{4}\mathscr D_B(\mathscr D^2f+2f)-\frac{3}{4}(\mathscr D_B\mathscr D_C\hat c_A^C-\mathscr D_A\mathscr D_C\hat c_B^C)\mathscr D^Bf\\ 
           &+\frac{1}{2}\left( \mathscr D_A\mathscr D_Bf-\frac{1}{2}\gamma_{AB}\mathscr D^2f \right)\mathscr D_C\hat c^{BC}+\frac{3}{8}\mathscr D_A(\hat c^{BC}\mathscr D_B\mathscr D_Cf)\\ 
           &-\frac{2\dot\varphi_1}{3\varphi_0}f\mathscr D_B\hat c^B_A+\frac{2\varphi_1}{3\varphi_0}\mathscr D_B(fN^B_A)-\frac{\varphi_1}{6\varphi_0}\hat c_A^B\mathscr D_B\psi\\ 
           &-\frac{2}{3}\mathscr D_B\left( \frac{\varphi_1}{\varphi_0}Y^B\mathscr D_C\hat c^C_A \right)-\frac{2\varphi_1}{3\varphi_0}(\mathscr D_AY_B)\mathscr D_C\hat c^{BC}\\
           &+\frac{2\omega+3}{8}\frac{\varphi_1\dot\varphi_1}{\varphi_0^2}\mathscr D_Af+\frac{2\varphi_1}{3\varphi_0}\mathscr D_A(\mathscr D^2f+2f)\\ 
           &-\frac{2\omega+3}{16}\left( \frac{\varphi_1}{\varphi_0} \right)^2\mathscr D_A\psi,
        \end{split}
    \end{equation}
\end{subequations}
where $\lie_Y\hat c_{AB}=Y^C\mathscr D_C\hat c_{AB}+\hat c_{CB}\mathscr D_AY^C+\hat c_{AC}\mathscr D_BY^C$ is the Lie derivative, and a similar expression for $\lie_Y\hat d_{AB}$ applies.
Thus, the news tensor $N_{AB}$ and the scalar $N$ change according to 
\begin{subequations}
    \label{eq-bms-ns}
    \begin{gather}
        \delta_\xi N_{AB}=f\dot N_{AB}+\lie_YN_{AB},\\ 
        \delta_\xi N=f\dot N+\psi N+Y^A\mathscr D_AN,
    \end{gather}
\end{subequations}
where $\lie_YN_{AB}$ has a similar expression to $\lie_Y\hat c_{AB}$.

Now, it is ready to decode the relation of memory effect, BMS symmetry and the degenerate vacua.


\section{Memory effect and the degenerate vacua}
\label{sec-mem}

As briefly introduced in section~\ref{sec-pols}, memory effect is the permanent change in the deviation vector between the initial and the final states without any GWs. 
These states are called nonradiative \cite{Flanagan:2015pxa}. 
In a nonradiative state, $N_{AB}=0$ and $N=0$.
Because of eqs.~\eqref{eq-bms-ns}, these states are invariant under any BMS transformation.
The vanishing $N_{AB}$ and $N$ does not imply either $\hat c_{AB}=0$ or $\varphi_1=0$.
If $\hat c_{AB}$ and $\varphi_1$ were zero, an arbitrary infinitesimal BMS transformation would generally result in nonvanishing $\hat c_{AB}$ and $\varphi_1$, referring to eqs.~\eqref{eq-bms-phi1} and \eqref{eq-bms-c}.
We are more interested in the special case where the initial and the final states are both vacua.
The definition of the vacuum in the scalar sector is simple, i.e., $N=0$.
But for the tensor sector, it is trickier.

To define a vacuum state in the tensor sector, one sets up a Newman-Penrose (NP) tetrad $\{l^a,n^a,m^a,\bar m^a\}$ at $\scri$ \cite{Newman:1961qr},
\begin{equation}
    \label{eq-np-tet}
    l^a=(\pd_r)^a,\quad n^a=-(\pd_u)^a+\frac{1}{2}(\pd_r)^a,\quad m^a=\frac{1}{\sqrt{2}r}\left[ (\pd_\theta)^a-i\csc\theta (\pd_\phi)^a \right],
\end{equation}
and $\bar m^a$ is the complex conjugate of $m^a$.
One can check that at $\scri$, they satisfy the following normalization conditions,
\begin{equation}
    \label{eq-np-norm}
    l^an_a=m^a\bar m_a=-1,
\end{equation}
and other contractions are zero, and so these vector are all null.
In addition, $l^a$ satisfies the geodesic equation.
Now, one can calculate some NP variables related to the outgoing tensor GW, given by \cite{Newman:1968uj}
\begin{subequations}
    \begin{gather}
        \Psi_4=C_{abcd}n^a\bar m^bn^c\bar m^d=-\frac{r}{2}\pd_u^2\hat c_{AB}\bar m^{A}\bar m^{B}+\cdots,\\ 
        \Psi_3=C_{abcd}\bar m^an^bl^cn^d=\frac{1}{2r}\bar m^A\mathscr D_BN^B_A+\cdots,
    \end{gather}
and
\begin{equation}
        \begin{split}\label{eq-impsi2}
        \Im\Psi_2=&\Im(C_{abcd}\bar m^an^bl^cm^d)\\ 
        =&\frac{1}{i8r}(-N_A^C\hat c_{BC}-\mathscr D_A\mathscr D_C\hat c^C_B+\mathscr D_B\mathscr D_C\hat c^C_A)(\bar m^Am^B-m^A\bar m^B)+\cdots,
        \end{split}
\end{equation}
\end{subequations}
where dots represent higher order terms, and $\Im$ is to take the imaginary part.
The vacuum state is the one in which the leading order terms in the above expressions vanish, together with $N_{AB}=0$, in GR \cite{Ashtekar:1981hw}.
Here, in BD, we also define the vacuum state in the same way.
So one has to solve for $\hat c_{AB}$ such that the right-hand side of the last equation vanishes at the leading order. 
For this purpose, it is easier to use the standard complex coordinate $\zeta=e^{i\phi}\cot\frac{\theta}{2}$ on the unit 2-sphere. 
In this coordinates, one has 
\begin{equation}
    \label{eq-m-com}
    m^a=-\sqrt{\frac{\bar\zeta}{2\zeta}}(1+\zeta\bar\zeta)(\pd_{\bar\zeta})^a.
\end{equation}
Therefore, the vanishing of the leading order term of the right-hand side of \eqref{eq-impsi2} is equivalent to
\begin{equation}
    \label{eq-vn-impsi2}
    \frac{(1+\zeta\bar\zeta)^4}{i16}(\mathscr D_{\bar\zeta}^2\hat c_{\zeta\zeta}-\mathscr D_\zeta^2\hat c_{\bar\zeta\bar\zeta})=0.
\end{equation}
The general solution to this equation is \cite{Strominger2014bms,Strominger:2014pwa}
\begin{equation}
    \label{eq-sol-vn-impsi2}
    \hat c_{\zeta\zeta}=-2\mathscr D_\zeta^2C(\zeta,\bar\zeta),
\end{equation}
for some function $C(\zeta,\bar\zeta)$ on the unit 2-sphere.
Thus, $C(\zeta,\bar\zeta)$ completely characterizes $\hat c_{\zeta\zeta}$ in vacuum state.
In summary, a vacuum is labeled by functions $C(\zeta,\bar\zeta)$ and $\varphi_1(\zeta,\bar\zeta)$, not by the leading order spacetime metric, i.e., the flat metric $\ud s^2_0=-\ud u^2-2\ud u\ud r+r^2\gamma_{AB}\ud x^A\ud x^B$.
This metric is shared by all vacua \cite{Geroch1977}.
As long as either $C(\zeta,\bar\zeta)$ or $\varphi_1(\zeta,\bar\zeta)$ changes, the vacuum changes.

This result has a profound implication. 
Let us perform an infinitesimal supertranslation with $f=\alpha(x^A)$ and $Y^A=0$ on the vacuum state, then $\hat c_{AB}$ transforms according to 
\begin{subequations}
\begin{equation}
    \label{eq-ctr}
    \delta_\alpha\hat c_{AB}=-2\mathscr D_A\mathscr D_B\alpha+\gamma_{AB}\mathscr D^2\alpha,
\end{equation}
or, in the complex coordinate system, 
\begin{equation}
    \label{eq-ctr-z}
    \delta_\alpha\hat c_{\zeta\zeta}=-2\mathscr D_\zeta^2\alpha.
\end{equation}
\end{subequations}
Therefore, the transformed $\hat c'_{\zeta\zeta}=-2\mathscr D_\zeta^2C'$ with $C'=C+\alpha$ is also a function on the unit 2-sphere.
This means that the state described by $\hat c'_{\zeta\zeta}$ is still vacuum, so there is no a unique vacuum in the tensor sector.
The degenerate vacua are related to each other via a supertranslation ($\alpha$).
However, an infinitesimal Lorentz transformation with $\alpha=0$ and $Y^A\ne0$ changes the vacuum state in the tensor sector, because 
\begin{equation}
    \label{eq-ctr-l}
    \delta_Y\hat c_{\zeta\zeta}=\lie_Y\hat c_{\zeta\zeta}-\frac{\psi}{2}\hat c_{\zeta\zeta},
\end{equation}
which does necessarily lead to $\hat c''_{\zeta\zeta}=-2\mathscr D_\zeta^2C''$ for some $C''(\zeta,\bar\zeta)$.
So the resultant state is no longer a vacuum.
Since we are now discussing the memory effect in BD, we should also study the transformation of $\varphi_1$. 
By eq.~\eqref{eq-bms-phi1}, an infinitesimal supertranslation does not change $\varphi_1$ in a vacuum state, but an infinitesimal Lorentz transformation generated by $Y^A$ does, 
\begin{equation}
    \label{eq-phi1-l}
    \delta_Y\varphi_1=\frac{\psi}{2}\varphi_1+Y^A\mathscr D_A\varphi_1.
\end{equation}
So this means that in the scalar sector, there are also degenerate vacua with respect to the Lorentz transformation, instead of the supertranslation.

Now, it is ready to discuss how memory effect is related to supertranslations and Lorentz transformations in BD. 
Consider an isolated system which is initially in a vacuum state before the retarded time $u_i$, with $\hat c_{\zeta\zeta}$ and $\varphi_1=\varphi_i$ constant in $u$.
Then, it starts to radiate GWs, and eventually settles down to a new vacuum state with $\hat c'_{\zeta\zeta}$ and $\varphi_1=\varphi_f$ after the time $u_f$. 
Now, begin with the memory effect for the tensor GW.
The above discussion shows that there exist two functions $C(\zeta,\bar\zeta)$ and $C'(\zeta,\bar\zeta)$ on the unit 2-sphere such that $\hat c_{\zeta\zeta}=-2\mathscr D_\zeta^2C$ and $\hat c'_{\zeta\zeta}=-2\mathscr D_\zeta^2C'$. 
The difference $\alpha=C'-C$ generates an infinitesimal supertranslation, which relates $\hat c'_{\zeta\zeta}$ to $\hat c_{\zeta\zeta}$ in the following way,
\begin{equation}
    \label{eq-ctr-3}
    \hat c'_{\zeta\zeta}-\hat c_{\zeta\zeta}=-2\mathscr D_\zeta^2\alpha.
\end{equation}
Therefore, the deviation vector changes permanently according to 
\begin{equation}
    \label{eq-dev-a}
    \Delta S_\zeta\approx-\frac{\mathscr D_\zeta^2\alpha}{r}S_0^\zeta+\order{\frac{1}{r^2}},
\end{equation}
due to the tensor GW.
This observation suggests the memory effect of the tensor GW is due to the supertranslation that induces the transition among the degenerate vacua in the tensor sector.

However, the supertranslation does not change $\varphi_1$ as discussed above, so the memory effect caused by the scalar GW is not generated by a supertranslation.
Rather, the Lorentz transformation does the work.
Suppose after the radiation, the initial scalar field $\varphi_1=\varphi_i$ changes to $\varphi'_1=\varphi_f$.
Then one may not be able to find a vector $Y^A$ such that $\delta_Y\varphi_1=\varphi_f-\varphi_i$, but a finite Lorentz transformation should relate $\varphi_i$ to $\varphi_f$.
Therefore, it is the Lorentz transformation that generates the transition among the vacua in the scalar sector and thus the memory effect of the scalar GW.

With the evolution equations~\eqref{eq-uu2-mdot} and \eqref{eq-uA2-ndot}, one can find the changes in $\hat c_{AB}$ and $\varphi_1$.
For the tensor memory effect, one integrates eq.~\eqref{eq-uu2-mdot} over $u$ from $u_i$ to $u_f$, and then, moves some terms to get 
\begin{equation}
    \label{eq-cc}
    \mathscr D_{\zeta}^2\Delta\hat c^{\zeta\zeta}=2\Delta m+\int_{u_i}^{u_f}\ud u\left[ \frac{8\pi G_0}{\varphi_0}\bar T_{uu}+\frac{1}{2}N_{\zeta\zeta}N^{\zeta\zeta}+\frac{2\omega+3}{2}\left( \frac{N}{\varphi_0} \right)^2 \right].
\end{equation} 
Here, eq.~\eqref{eq-vn-impsi2} has been used. 
With eq.~\eqref{eq-ctr-3}, one determines \cite{Strominger:2014pwa}
\begin{equation}
    \label{eq-sol-al}
    \begin{split}
    \alpha(\zeta,\bar\zeta)=&-\int\ud\zeta'\ud\bar\zeta'\gamma_{\zeta'\bar\zeta'}G(\zeta,\bar\zeta;\zeta',\bar\zeta')\Bigg\{ \Delta m+\int_{u_i}^{u_f}\ud u\Bigg[ \frac{4\pi G_0}{\varphi_0}\bar T_{uu}\\ 
    &+\frac{1}{4}N_{\zeta\zeta}N^{\zeta\zeta}+\frac{2\omega+3}{4}\left( \frac{N}{\varphi_0} \right)^2 \Bigg]\Bigg\},
    \end{split}
\end{equation}
where $G(\zeta,\bar\zeta;\zeta',\bar\zeta')$ is the Green's function, given by 
\begin{equation*}
    \label{eq-def-g}
    G(\zeta,\bar\zeta;\zeta',\bar\zeta')=-\frac{1}{\pi}\sin^2\frac{\Theta}{2}\ln\sin^2\frac{\Theta}{2},\quad \sin^2\frac{\Theta}{2}\equiv\frac{|\zeta-\zeta'|^2}{(1+\zeta\bar\zeta)(1+\zeta'\bar\zeta')},
\end{equation*}
such that $\mathscr D_\zeta^2\mathscr D_{\bar\zeta}^2G(\zeta,\bar\zeta;\zeta',\bar\zeta')=-\gamma_{\zeta\bar\zeta}\delta^2(\zeta-\zeta')+\cdots$.
By eq.~\eqref{eq-sol-al}, one concludes that the tensor memory effect is due to the transition among the tensor vacua caused by the null energy fluxes passing through $\scri$, exactly the same as in GR \cite{Strominger:2014pwa}.
The null energy fluxes include the matter, the tensor GW and the scalar GW fluxes, as displayed in the square brackets in sequence.
In terms of the terminology of ref.~\cite{Bieri:2013ada}, the first term in the curly brackets of eq.~\eqref{eq-sol-al} is the \emph{ordinary memory}, and the remaining ones represent the \emph{null memory}.
Of course, the null memory caused by the scalar GW is new, which is called T memory in ref.~\cite{Du:2016hww}.

For the scalar memory effect, contract both sides of eq.~\eqref{eq-uA2-ndot} with $\mathscr D^A$, integrate over $u$ from $u_i$ to $u_f$, and act $\mathscr D^{-2}$ on both sides.
Then, one obtains that 
\begin{equation}
    \label{eq-sol-p}
    \begin{split}
    \frac{2\omega+3}{16\varphi_0^2}\Delta\varphi_1^2=&-\frac{\pi G_0}{\varphi_0}\Delta\bar T_{rr}+\frac{\Delta(\hat c_A^B\hat c_B^A)}{32}-\int_{u_i}^{u_f}m\ud u+\frac{1}{2\pi}\int\ud\zeta'\ud\bar\zeta'\gamma_{\zeta'\bar\zeta'}\mathcal G(\zeta,\bar\zeta;\zeta',\bar\zeta')\\
    &\mathscr D^A\Bigg[ \Delta N_A+\int_{u_i}^{u_f}\ud u\left( \frac{8\pi G_0}{\varphi_0}\bar T_{uA}-\frac{1}{4}N_C^B\mathscr D_A\hat c^C_B+\frac{2\omega+3}{2\varphi_0^2}N\mathscr D_A\varphi_1 \right) \Bigg],
    \end{split}
\end{equation}
where $\mathcal G(\zeta,\bar\zeta;\zeta',\bar\zeta')=\ln\sin^2\frac{\Theta}{2}$ is another Green's function satisfying $\pd_\zeta\pd_{\bar\zeta}\mathcal G(\zeta,\bar\zeta;\zeta',\bar\zeta')=2\pi\delta^2(\zeta-\zeta')-\gamma_{\zeta\bar\zeta}/2$ \cite{Pasterski:2015tva},
and if the terms in the square brackets were written in the $\zeta$ coordinate system, the expression would be very clumsy.
So one can consider an experimentally relevant process whose initial state is some vacuum state with an initial $\varphi_i$ (and $\hat c_{\zeta\zeta}$), and the final state contains a single stationary back hole. 
Due to the no-hair theorem, the final $\varphi_f=0$ \cite{Hawking1972bhbd,Sotiriou:2012nh}. 
Therefore, the left-hand side of eq.~\eqref{eq-sol-p} is proportional to $\Delta\varphi_i^2$.
The terms in the round brackets in the second line are the angular momentum fluxes of the matter, the tensor GW and the scalar GW, respectively \cite{Hou:2017cjy}.
This means that the scalar memory effect is related to the transition among the scalar vacua induced by the angular momentum fluxes through $\scri$.
Again, in the terminology of ref.~\cite{Bieri:2013ada}, one might call the effect corresponding to the round brackets the null scalar memory, and the one to the remaining terms the ordinary scalar memory.

As a final remark, we would like to briefly discuss the spin and the CM memory effects in BD. 
By the arguments in refs.~\cite{Pasterski:2015tva,Nichols:2018qac}, these effects are due to the leading order term of $g_{uA}$ component. 
According to eq.~\eqref{eq-met-uA}, the leading order term of $g_{uA}$ contains no contributions from the scalar field $\varphi$.
So in tensor sector in BD, there are also similar spin and CM memory effects  to those in GR, but in scalar sector, neither  spin memory nor CM memory exists. 
Of course, this does not mean that $\varphi$ plays no role in the spin or CM memory. 
In fact, its angular momentum flux also generates the spin and CM memories, just like other angular momentum fluxes \cite{Pasterski:2015tva,Nichols:2018qac}.
Note that the spin memory effect is induced by the curl part of the angular momentum fluxes as clearly demonstrated by eq.~(4.10) in ref.~\cite{Flanagan:2015pxa}, while the scalar displacement memory effect is due to the divergence part of the angular momentum fluxes, referring to eq.~\eqref{eq-sol-p}.
The relation between the CM memory effect and the angular momentum fluxes are more complicated, so its discussion is presented in ref.~\cite{Hou:2020wbo}.

\section{Conclusion}
\label{sec-con}

From the above discussion, one understands that the memory effect of an isolated system in BD shares some similarities with that in GR and also has its own specialties. 
In both theories, there are memory effects for the tensor GW, which are induced by the passage of the null energy fluxes through $\scri$.
The supertranslations transform the degenerate vacua in the tensor sector, which makes the tensor memory effect possible.
However, in BD, there exists the scalar GW.
It not only contributes to the tensor memory effect by providing a new energy flux, but also has its own memory effect. 
This new memory effect is due to the angular momentum fluxes penetrating $\scri$.
There are also degenerate vacua in the scalar sector, which are transformed into each other via the Lorentz transformations.

\acknowledgments
This work was supported by the National Natural Science Foundation of China under grant Nos.~11633001 and 11920101003, and the Strategic Priority Research Program of the Chinese Academy of Sciences, grant No.~XDB23000000.
SH was also supported by Project funded by China Postdoctoral Science Foundation (No.~2020M672400).

\appendix

\section{Memories and soft theorems}
\label{sec-app-mst}

In this appendix, we will briefly discuss the relation between memory effects and soft theorems.
For this purpose, we will first compute the memory effects in the scattering of large massive objects in subsection~\ref{sec-app-bm}.
Then, subsection~\ref{sec-app-st} focuses on the soft theorems.
In this section, the abstract index notation will not be used.

\subsection{Burst memory waves}
\label{sec-app-bm}

Consider the scattering of large massive objects, such as stars and black holes. 
The scattering generates GWs that may produce the displacement memory effects at the null infinity. 
For the purpose of calculating the GW waveforms, we can linearize eqs.~\eqref{eq-ein} and \eqref{eq-s} by setting $g_{\mu\nu}=\eta_{\mu\nu}+h_{\mu\nu}$ and $\varphi=\varphi_0+\chi$ with $\eta_{\mu\nu}$ the flat metric, and $h_{\mu\nu}$ and $\chi$ small.
Further, one replaces $h_{\mu\nu}$ by $\tilde h_{\mu\nu}=h_{\mu\nu}-\eta_{\mu\nu}\eta^{\rho\sigma}h_{\rho\sigma}/2-\eta_{\mu\nu}\chi/\varphi_0$, and imposes the Lorenz gauge condition ($\pd^\nu\tilde h_{\mu\nu}=0$) to obtain \cite{Liang:2017ahj,Hou:2017cjy},
\begin{subequations}
    \begin{gather}
        \Box\tilde h_{\mu\nu}=-\frac{16\pi G_0}{\varphi_0}T_{\mu\nu},\\ 
        \Box\chi=\frac{8\pi G_0}{2\omega+3}T,
    \end{gather}
\end{subequations}
where $\Box=\pd_\mu\pd^\mu$.
According to ref.~\cite{Kovacs:1977uw}, the matter stress-energy tensor can be taken as
\begin{equation}
    \label{eq-mst}
    T_{\mu\nu}=\sum_j\int m^{(j)}u_\mu^{(j)}u_\nu^{(j)}\delta^{(4)}[x-X(\tau^{(j)})]\ud\tau^{(j)},
\end{equation}
which is suitable for a systems of stars of masses $m^{(j)}$, moving at the 4-velocities $u^a_{(j)}$ along the trajectories $X(\tau^{(j)})$ parameterized by the proper times $\tau^{(j)}$.
The index $j$ labels different stars, and can be freely raised or lowered.
One can use the method of Green's function to obtain the wave solutions, and the Green's function is taken to be the flat space propagator \cite{Thorne1975wfs},
\begin{equation}
    \label{eq-gf}
    {}_0G(x,y)=\frac{1}{4\pi}\delta_\text{ret}\left[ \frac{1}{2}\eta_{\mu\nu}(x^\mu-y^\mu)(x^\nu-y^\nu) \right],
\end{equation}
in which $\delta_\text{ret}$ vanishes if $y^\mu$ is in the causal past of $x^\mu$, and otherwise, it is the Dirac delta function.
With these, one can easily calculate to get,
\begin{subequations}
    \begin{gather}
        \tilde h_{\mu\nu}=\frac{4G_0}{\varphi_0r}\sum_j\frac{p_\mu^{(j)}p_\nu^{(j)}}{k\cdot p^{(j)}},\\ 
        \chi=\frac{2G_0}{(2\omega+3)r}\sum_j\frac{m_{(j)}^2}{k\cdot p^{(j)}},\label{eq-chi-v}
    \end{gather}
\end{subequations}
where $r$ is the distance between the observer and the collision region, $p^{(j)}_\mu=m^{(j)}u_\mu^{(j)}$, $k^\mu=(-1,-\hat n)$ with $\hat n$ a unit spatial vector pointing from the source to the observer, and $k\cdot p^{(j)}=k^\mu p_\mu^{(j)}$. 
It seems that eq.~\eqref{eq-chi-v} is negative, but since  $k^\mu$ is pointing to the past,  $k\cdot p^{(j)}>0$, 
so eq.~\eqref{eq-chi-v} is positive.

Although $\tilde h_{\mu\nu}$ differs from $h_{\mu\nu}$, ref.~\cite{Hou:2017bqj} has shown that the plus and the cross polarizations are represented by the transverse-traceless (TT) part of $\tilde h_{\mu\nu}$, so the displacement memory effect in the tensor sector is given by 
\begin{equation}
    \label{eq-dm-t}
    \Delta h_{\mu\nu}^\text{TT}(k)=\frac{4G_0}{\varphi_0r}\left(\sum_{j=1}^{\mathcal O}\frac{p_\mu^{'(j)}p_\nu^{'(j)}}{k\cdot p^{'(j)}} - \sum_{j=1}^{\mathcal I}\frac{p_\mu^{(j)}p_\nu^{(j)}}{k\cdot p^{(j)}}\right)^\text{TT},
\end{equation}
where $p^{(j)}$ are the momenta of the incoming stars, $p^{'(j)}$ are for the outgoing stars, and $\mathcal I$ and $\mathcal O$ are the numbers of the incoming and the outgoing stars, respectively.
The displacement memory effect in the scalar sector is 
\begin{equation}
    \label{eq-dm-s}
    \Delta\chi(k)=\frac{2G_0}{(2\omega+3)r}\left(\sum_{j=1}^{\mathcal O}\frac{m_{(j)}^2}{k\cdot p^{'(j)}}-\sum_{j=1}^{\mathcal I}\frac{m_{(j)}^2}{k\cdot p^{(j)}}\right).
\end{equation}
Since $\chi$ represents the leading order correction to $\varphi$, $\Delta\chi=\Delta\varphi_1/r$, and the above expression indeed describes the displacement memory in the scalar sector.

By eq.~\eqref{eq-dm-t}, the tensor memory in BD takes exactly the same form as that in GR, modulo some factor difference \cite{Strominger:2014pwa}. 
As discussed in the next section, the action for the tensor GW, accurate up to an appropriate order, is also the same in both theories.  
Therefore, the memory effect and the soft theorem are correctly related to each other in BD.
So the next section focus on the soft theorem for the scalar GW.

\subsection{Soft theorems}
\label{sec-app-st}

In this subsection, the soft theorem is derived and generalized to BD theory. 
Here, the matter action is simply taken to be \cite{He:2014laa} 
\begin{equation}
    \label{eq-mact}
    S_m=\int\ud^4x\sqrt{-g}\left( -\frac{1}{2}\nabla_\mu\Phi\nabla^\mu\Phi-\frac{1}{2}m^2\Phi^2 \right),
\end{equation}
where $\Phi$ is a scalar field with mass $m$.
Now, perturb the total action \eqref{eq-act-bd} to suitable order in $h_{\mu\nu}$ and $\chi$.
Then, one obtains an action which contains terms of the form $\pd\chi\pd h$.
In order to remove them, one defines $h_{\mu\nu}=\bar h_{\mu\nu}-\eta_{\mu\nu}\chi/\varphi_0$, so one obtains a Lagrange whose kinetic terms are 
\begin{subequations}
    \begin{equation}
        \label{eq-l-kin}
        \begin{split}
            \mathcal L_\text{kin}=&-\frac{1}{2}\pd_\mu\Phi\pd^\mu\Phi-\frac{1}{2}m^2\Phi^2-\frac{2\omega+3}{16\pi G_0\varphi_0}\pd_\mu\chi\pd^\mu\chi\\
            &-\frac{\varphi_0}{16\pi G_0}\left( \frac{1}{2}\pd_\rho\bar h_{\mu\nu}\pd^\rho\bar h^{\mu\nu} -\pd^\rho\bar h_{\mu\rho}\pd_\sigma \bar h^{\mu\sigma}+\pd^\mu\bar h\pd^\nu\bar h_{\mu\nu}-\frac{1}{2}\pd_\mu\bar h\pd^\mu\bar h\right),
        \end{split}
    \end{equation}
and the interaction terms relevant for our purpose are 
\begin{equation}
    \label{eq-l-int}
    \mathcal L_\text{int}=\frac{\chi}{2\varphi_0}\left(\pd_\mu\Phi\pd^\mu\Phi+m^2\Phi^2\right)
    +\frac{\bar h^{\mu\nu}}{2}\left[ \pd_\mu\Phi\pd_\nu\Phi-\frac{\eta_{\mu\nu}}{2}\left(\pd_\rho\Phi\pd^\rho\Phi +m^2\Phi^2\right)\right].
\end{equation}
\end{subequations}
The interaction for $\chi$ differs from the one in ref.~\cite{Campiglia:2017dpg}.
From these results, one knows that $\bar h_{\mu\nu}$ behaves just like the one in GR, if one ignores the mass term for $\Phi$.
As argued in ref.~\cite{He:2014laa}, the mass term does not affect the soft theorem for the spin-2 graviton, so ref.~\cite{He:2014laa}'s results in section~4 can be used by suitably rescaling their eq.~(4.7).
This means that the memory effect in the tensor sector is associated with the soft (spin-2) graviton theorem via the Fourier transformation in the tensor sector in BD.
In fact, with the arguments in ref.~\cite{Strominger2014bms}, the infrared triangle also exists in the tensor sector in BD.

So now, we will repeat the computation in section~4 of ref.~\cite{He:2014laa} for the scalar field $\chi$.
For that purpose, we first rescale $\chi$, that is, we consider the canonical scalar field $\hat\chi=Z\chi$ with $Z=\sqrt{\frac{2\omega+3}{16\pi G_0\varphi_0}}$.
The Feynman rule for the interaction $\frac{\hat\chi}{2Z\varphi_0}(\pd_\mu\Phi\pd^\mu\Phi+m^2\Phi^2)$ is 
\begin{equation}
    \label{eq-fr}
\begin{fmffile}{diagram}
    \parbox{80pt}{
\begin{fmfgraph*}(80,60)
    \fmfpen{thin}
    \fmfleft{i1}
    \fmfright{o1,o2}
    \fmf{fermion,label=$p_1$,label.side=left}{i1,v1}
    \fmf{dashes}{v1,o1}
    \fmf{fermion,label=$p_2$,label.side=right}{v1,o2}
\end{fmfgraph*}
    }\quad =\frac{i}{Z\varphi_0}\left(p_1\cdot p_2+m^2\right).
\end{fmffile}
\end{equation}
Now, consider a process with $\mathcal I$ incoming and $\mathcal O$ outgoing scalars $\Phi$, whose scattering amplitude is denoted as $\mathcal M(p'_1,\cdots,p'_{\mathcal O};p_1,\cdots,p_{\mathcal I})$, and a similar process with one extra soft $\chi$ emitted, whose amplitude is denoted as $\mathcal M(q,p'_1,\cdots,p'_{\mathcal O};p_1,\cdots,p_{\mathcal I})$.
The contributing Feynman diagrams to the second amplitude includes the following typical one,

\begin{equation}
    \label{eq-ma}
\begin{fmffile}{diagram2}
    \parbox{60pt}{
   \begin{fmfgraph*}(60,60)
       \fmfpen{thin}
       \fmfleft{i1,i2,i3}
       \fmfright{o1,o2,os,o3,o4}
       \fmf{fermion}{i1,v}
       \fmfblob{100}{v}
       \fmf{fermion}{i2,v}
       \fmf{fermion}{i3,v}
       \fmflabel{$p_1$}{i1}
       \fmflabel{$\vdots$}{i2}
       \fmflabel{$p_{\mathcal I}$}{i3}
       \fmf{fermion}{v,o1}
       \fmf{fermion}{v,v1,o2}
       \fmf{dashes}{v1,os}
       \fmf{fermion}{v,o3}
       \fmf{fermion}{v,o4}
       \fmflabel{$p'_1$}{o1}
       \fmflabel{$p'_j$}{o2}
       \fmflabel{$q$}{os}
       \fmflabel{$\vdots$}{o3}
       \fmflabel{$p'_{\mathcal O}$}{o4}
   \end{fmfgraph*}
   }
   \quad\quad =\frac{1}{Z\varphi_0}\frac{p'_j\cdot(p'_j-q)+m^2}{-(p'_j-q)^2-m^2}\mathcal M(p'_1,\cdots,p'_j-q,\cdots,p'_{\mathcal O};p_1,\cdots,p_{\mathcal I}).
\end{fmffile}\vspace{5mm}
\end{equation}
By moving the dashed line to other outgoing external legs, one obtains the remaining diagrams whose amplitudes are similar to the above one, and if the dashed line is attached to one of the incoming external legs, one has to add a minus sign. 
Note the factor becomes
\begin{equation}
    \label{eq-ma-sf}
\frac{p'_j\cdot(p'_j-q)+m^2}{-(p'_j-q)^2-m^2}=\frac{-p_j\cdot q}{2p_j\cdot q}\rightarrow-\frac{1}{2},
\end{equation}
exactly.
From this result, one knows that this soft theorem cannot be the Fourier transform of eq.~\eqref{eq-dm-s}.
This is consistent with the fact that under the supertranslation, $\varphi_1\propto\chi$ does not change if it is in the vacuum state. 

As is known, the spin memory effect in GR is related to the subleading soft theorem \cite{Cachazo:2014fwa,Pasterski:2015tva}, and the angular momentum flux arriving at $\scri$ induces it \cite{Flanagan:2015pxa}. 
By eq.~\eqref{eq-sol-p}, the displacement memory effect in the scalar sector also has something to do with the angular momentum flux, so one expects some new form of the subleading soft theorem exists for the scalar memory effect. 
This deserves further studies, and will be discussed in the future.

\bibliographystyle{JHEP}
\bibliography{MemoryHorndeski.bbl}

\end{document}